\documentclass[twocolumn, superscriptaddress, amsmath,amssymb,aps]{revtex4-2}

\usepackage{bm}
\usepackage{graphicx}
\usepackage{amsmath, amsthm}
\usepackage{hyperref}
\usepackage{tikz}
\usepackage{multirow}
\usepackage{xcolor}

\begin{document}

\title{Mechanical signaling cascades}

\author{Michelle Berry}
\affiliation{Department of Physics, Syracuse University, Syracuse, NY}

\author{Yongjae Kim}
\affiliation{Department of Polymer Science and Engineering, University of Massachusetts, Amherst, MA}

\author{David Limberg}
\affiliation{Department of Polymer Science and Engineering, University of Massachusetts, Amherst, MA}

\author{Ryan C. Hayward}
\affiliation{Department of Chemical and Biological Engineering, University of Colorado, Boulder, CO}

\author{Christian D. Santangelo}
\affiliation{Department of Physics, Syracuse University, Syracuse, NY}

\date{\today}

\begin{abstract}
Mechanical computing has seen resurgent interest recently owing to the potential to embed sensing and computation into new classes of programmable metamaterials. To realize this, however, one must push signals from one part of a device to another, and do so in a way that can be reset robustly. We investigate the propagation of signals in a bistable mechanical cascade uphill in energy. By identifying a penetration length for perturbations, we show that signals can propagate uphill for finite distances and map out parameters for this to occur. Experiments on soft elastomers corroborate our results.
\end{abstract}

\maketitle

Mechanical devices that compute have apparently existed for thousands of years \cite{chase1980history}, yet have fallen out of favor since the advent of the modern electronic computer. However, the recent interest in mechanical metamaterials has led to a resurgence of interest in new forms of mechanical computing using modern -- and predominantly soft -- materials \cite{bertoldi2017flexible,yasuda2021mechanical}, resulting in demonstrations of stable memory \cite{chen2021reprogrammable}, boolean logic \cite{ion2017digital,jiang2019bifurcation,mei2021mechanical, meng2021bistability,song2019additively, waheed2020boolean, merkle2018mechanical}, and pattern recognition \cite{fang2016pattern} among other computational and logical tasks. The dream is to create materials that change their behavior based on simple logic in response to external stimuli. To achieve that, in addition to logical elements, one requires convenient methods to transmit signals from one part of a material to another and a way to reset the state of a device or signal.

Recent work has demonstrated devices that can propagate mechanical signals by taking advantage of a series of interacting bistable elements of varying design \cite{deng2021nonlinear, yasuda2020transition, hwang2018input, hwang2021topological, katz2018solitary, katz2019solitary, nadkarni2014dynamics, nadkarni2016universal}. Previous work has shown that for asymmetric bistable units, transitioning from a higher to lower energy state allows for propagation over arbitrarily long distances \cite{deng2019propagation, deng2020nonlinear, nadkarni2016unidirectional, raney2016stable}. When utilizing symmetric bistable units, a decreasing grading of the interaction energy between bistable elements \cite{hwang2018solitary} or the energy barrier of the element itself \cite{librandi2021programming} allows for stable propagation. Specifically designed bistable elements \cite{librandi2021programming} or active components \cite{browning2019reversible} can produce reversible signal propagation. 

In this paper, we focus on a horizontal chain of bistable units (Fig. \ref{unitcell}(b)) which act as relays that can each be in either a ``left'' (no signal) or ``right'' (signal) state \cite{raney2016stable, nadkarni2014dynamics}. In order to perform multiple computations using chains with these bistable elements, we will need the ability to reset our system by sending a signal in both directions along the chain, indicating that it is important to also understand whether signals can propagate along uniform chains with no bias or a bias infavor of the ``left'' state. 

To further explore this signaling cascade and gain insight into how many mechanical computations can really be performed using coupled, bistable elements, we consider this problem for finite chains of bistable elements as a function of both the difference in minimum energies, $\Delta$, and barrier height $E$ (Fig. \ref{unitcell}(b)). Specifically, we look at the cases where the potential is symmetric and favors neither state, and cases where the potential favors the ``left'' state. Since we imagine working primarily with soft materials, we focus on the overdamped limit. This provides us with a set of simple scaling laws governing when signals propagate in terms of the number of elements and the energy of the bistable elements and provides limits on the number of computations that could be conceivably performed and reset.

\section{Mechanical Signalling Cascades}

\subsection{Bistable elements}

\begin{figure}
    \centering
    \includegraphics[width=\columnwidth]{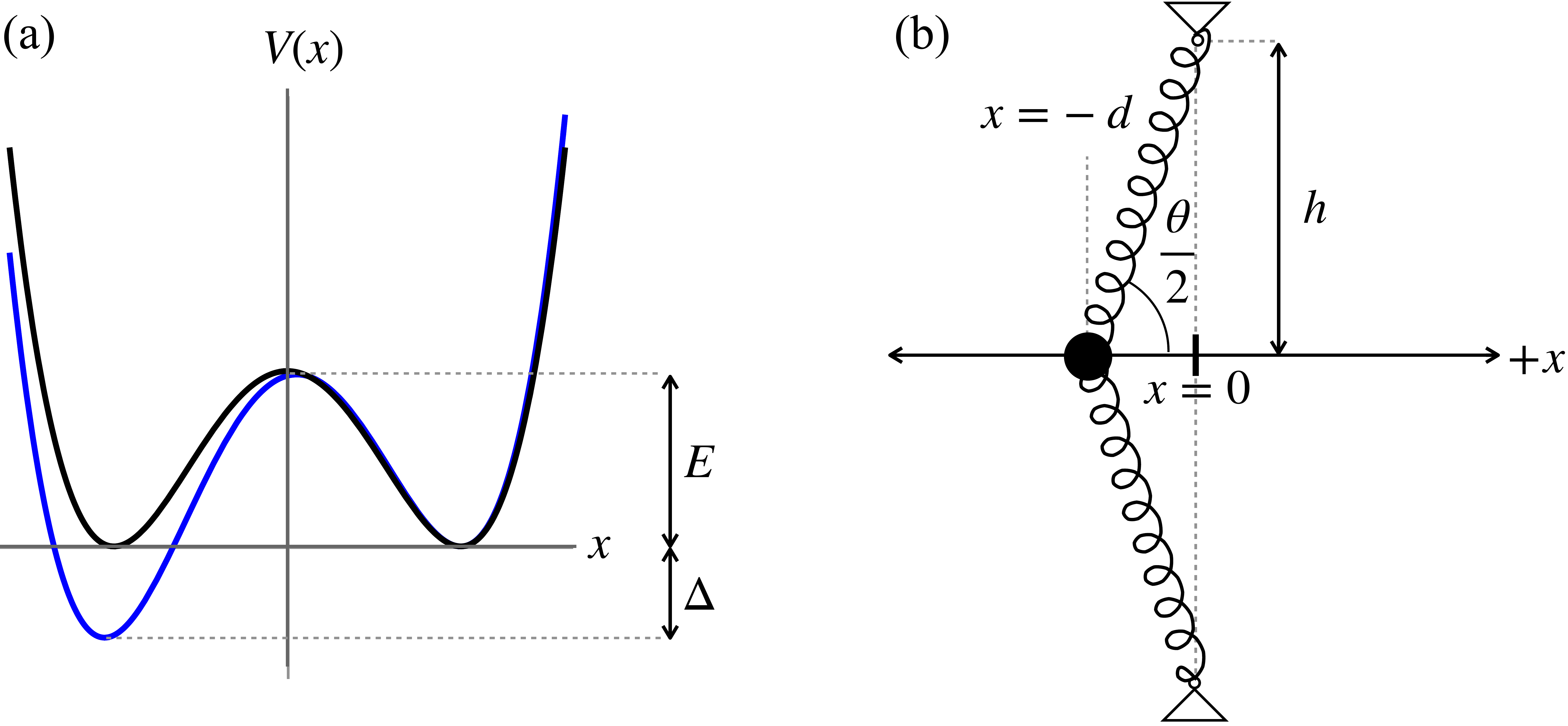}
    \caption{(a) The general shape of a bistable potential. The black line represents a symmetric potential with barrier height $E$. The blue line represents an asymmetric potential biased to the left where $\Delta$ is the potential difference between the two states. We measure the barrier height $E$ with respect to the higher minimum. (b)  An elastic bistable unit at equilibrium. The linear springs each have stiffness $k$. The torsional spring located at the point mass $m$ (filled black circle) has torsional modulus $s$ and equilibrium angle $\theta_0$. The potential minima for the bistable unit are located at $x=\pm d$. The potential barrier is located at $x=0$.}
    \label{unitcell}
\end{figure}

We consider a series of mechanical relays whose state is represented by a scalar variable $x$ subject to a bistable potential, $V(x)$ (Fig. \ref{unitcell}(a)). The potential is characterized by two minima having a difference in energy, $\Delta$. The energy barrier between them is given by $E$ and measures the barrier height with respect to the higher energy minimum. Though we will pursue a theoretical approach that is agnostic on the detailed form of the bistable elements, it is helpful to have a particular mechanical model in mind that can be implemented both experimentally and in simulations. We will model bistable elements as two linear springs (Fig. \ref{unitcell}(b)) with ends attached to fixed points \cite{nadkarni2014dynamics}. 

The potential energy of this two-spring bistable element is 
\begin{equation}\label{eq:beampotential}
    V(x) = b\left( \sqrt{x^2+h^2}-\sqrt{h^2+d^2} \right)^2
\end{equation}
where $x$ is the horizontal displacement of the point mass, $2 h$ is the distance between fixed vertices, and $\sqrt{h^2+d^2}$ is the equilibrium length of the two springs. The two minima are at $x= \pm d$, and the energy barrier height is $E = b [\sqrt{h^2+d^2} - h]^2$. To bias the bistable element towards one of the two stable states, we can add a torsional spring at the mass. This adds an additional term of the form
\begin{equation}\label{eq:torsionalpotential}
    V_{tor}(\theta) = \frac{1}{2} s \left(\theta-\theta_0 \right)^2
\end{equation}
to the potential energy of the bistable element, where $s$ is the torsional modulus, $\theta_0$ is the equilibrium angle of that modulus, and $\tan(\theta/2) = h/x$. 

Since we are interested primarily in signals driven up a potential barrier, we will choose the equilibrium angle to yield zero torsional energy when the bistable element is in the left-most state. This leads to an approximate barrier height,
\begin{equation}
    E = b \left( h - \sqrt{h^2+d^2} \right)^2 + \frac{s}{2}\left(\pi - \theta_0\right)^2 + \mathcal{O}(t^2),
\end{equation}
and asymmetry,
\begin{equation}
     \Delta = \frac{s}{2} \left( 2 \pi - 4 \tan^{-1} \frac{h}{d} \right)^2 + \mathcal{O}(t^2).
\end{equation}
Thus, $b$ predominantly controls the barrier height $E$ while $s$ controls the energy difference between minima, $\Delta$ (Fig. \ref{unitcell}(a)).

\subsection{Equations of Motion}

\begin{figure}
    \centering
    \includegraphics[width=\columnwidth]{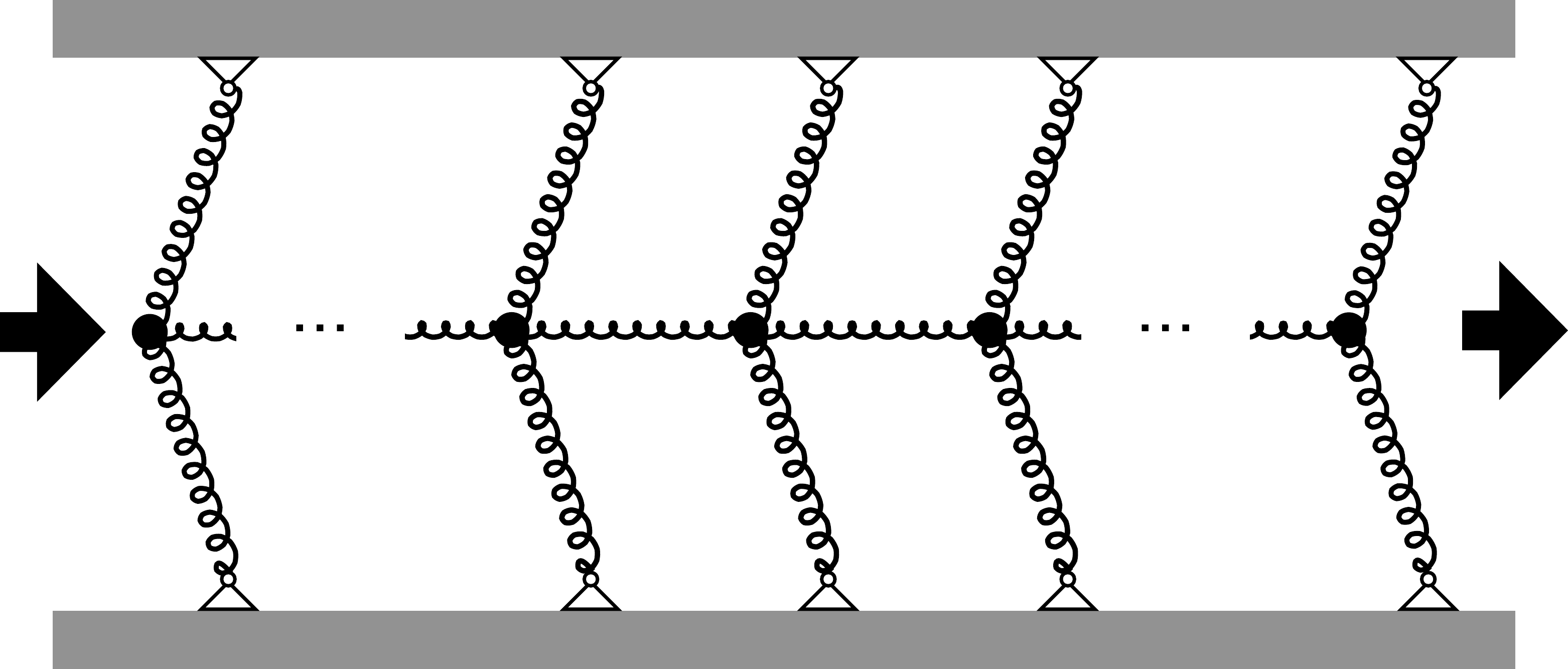}
    \caption{A finite length wire in the ``left'' (no signal) state. Each bistable element is at the $x=-d$ potential minimum. When the left-most point mass is pushed over the energy barrier and into the $x=+d$ energy minimum, the interaction spring connecting adjacent point masses allow that transition to propagate along the wire. If the right-most mass moves to the right, the entire wire is in the ``right'' (signal) state, and we say that the signal fully propagated along the wire. If only a portion of the bistable elements transition to the $x=+d$ energy minimum, we say that the signal only propagated a finite distance. }
    \label{chain}
\end{figure}

To form the one dimensional chain of bistable elements, we connect neighboring elements together at the point masses with linear springs of rest length $a$ and spring constant $k$ (Fig. \ref{chain}) \cite{nadkarni2014dynamics}. The position of the $n$-th point mass with respect to the fixed end points of the beam is given by $x_n$. Writing out Newton's second law for the $i$th mass gives
\begin{equation}\label{eq:EOM}
    m\frac{d^2x_i}{dt^2}-k[x_{i+1}-2x_i+x_{i-1}]+\gamma\frac{dx_i}{dt}+\frac{dV(x_i)}{dx_i}=0
\end{equation}
where $\gamma$ is a friction coefficient. To obtain the continuum limit, we make the variable change $x_{i\pm n}=u(y\pm na,t)$. The function $u(y,t)$ now represents the displacement of the beam at location $y$ along the wire. Taking the limit where $a\to 0$, the equation of motion for the $i$th mass becomes
\begin{equation}
    m\frac{\partial^2u}{\partial t^2}-ka^2\frac{\partial^2u}{\partial y^2}+\gamma\frac{\partial u}{\partial t}+\frac{dV}{du}=0.
\end{equation}  

Before we analyze the equation of motion, we replace $y$, $t$, and $u$ with dimensionless variables. First, we rescale $u\rightarrow a\tilde{u}$, $y\rightarrow a\tilde{y}$, and $t\rightarrow \tau\tilde{t}$ where $a$ is the length of the interaction springs and $\tau$ is some characteristic time.
\begin{equation}
    \frac{ma}{\tau^2}\frac{\partial^2\tilde{u}}{\partial \tilde{t}^2}-ka\frac{\partial^2\tilde{u}}{\partial \tilde{y}^2}+\frac{\gamma a}{\tau}\frac{\partial \tilde{u}}{\partial \tilde{t}}+\frac{1}{a}\frac{dV}{d\tilde{u}}=0
\end{equation}
Rescaling $V\rightarrow E V'$ by the barrier height of the potential, $E$, as defined in Fig. (\ref{unitcell}), we obtain
\begin{equation}
    \frac{ma}{\tau^2}\frac{\partial^2\tilde{u}}{\partial \tilde{t}^2}-ka\frac{\partial^2\tilde{u}}{\partial \tilde{y}^2}+\frac{\gamma a}{\tau}\frac{\partial \tilde{u}}{\partial \tilde{t}}+\frac{E}{a}\frac{dV'}{d\tilde{u}}=0
\end{equation}
Finally, we multiply the entire equation by the ratio $a/E$.
\begin{equation}\label{eq:newtoneq}
    \frac{ma^2}{E\tau^2}\frac{\partial^2\tilde{u}}{\partial \tilde{t}^2}-\frac{ka^2}{E}\frac{\partial^2\tilde{u}}{\partial \tilde{y}^2}+\frac{\gamma a^2}{E\tau}\frac{\partial \tilde{u}}{\partial \tilde{t}}+\frac{dV'}{d\tilde{u}}=0
\end{equation}

All terms in the equation of motion for the $n$th mass are now dimensionless. When ${ma^2}/{E\tau^2} \ll 1$, the system is overdamped and the first term of Eq. (\ref{eq:newtoneq}) can be neglected. To reduce the number of coefficients in Eq. (\ref{eq:newtoneq}), we rescale $\tilde{y}$ and $\tilde{t}$ again in the following way:
\begin{equation}
    y'=\sqrt{\frac{E}{ka^2}}\tilde{y}\;\;\;\;\;\;\;t'=\frac{E\tau}{\gamma a^2}\tilde{t}
\end{equation}
The equation of motion in the overdamped limit is now
\begin{equation}\label{eq:overdamped}
    -\frac{\partial^2\tilde{u}}{\partial y'^2}+\frac{\partial \tilde{u}}{\partial t'}+\frac{\partial V'}{\partial \tilde{u}}=0
\end{equation}
with all dimensionless terms and variables. We have the following relationships between our physical variables and the new dimensionless variables: 
\begin{equation}
    \begin{aligned}
        y'&=\sqrt{\frac{E}{ka^2}}\tilde{y}=\sqrt{\frac{E}{ka^2}}\frac{1}{a}y=\frac{1}{a^2}\sqrt{\frac{E}{k}}y\\
        t'&=\frac{E\tau}{\gamma a^2}\tilde{t}=\frac{E\tau}{\gamma a^2}\frac{1}{\tau}t=\frac{E}{\gamma a^2} t
    \end{aligned}
\end{equation}
Thus, given a solution to Eq. (\ref{eq:overdamped}), $\tilde{u}(y',t')$, we have
\begin{equation}\label{eq:soln}
    u(y,t) = a \tilde{u}\left( \sqrt{\frac{E}{k}} \frac{y}{a^2}, \frac{E}{\gamma a^2} t \right).
\end{equation}

Finally, if the last element of the chain is free, this requires $\partial u/\partial y |_{y=L} = 0$ on the rightmost boundary at $y=L$. Therefore, $\partial \tilde{u}/\partial y' = 0$ when $y' = L' = \sqrt{E/k} L/a^2$. On the left-most boundary at $y=y'=0$, we fix the initial displacement, $\tilde{u}(0)$.

\subsection{Solutions}
If we look for stationary solutions, we can find a first-integral for Eq. (\ref{eq:overdamped}) resulting in
\begin{equation}\label{eq:firstint}
    -\frac{1}{2} \left( \frac{\partial \tilde{u}}{\partial y'} \right)^2 + V'( \tilde{u}(y') ) = C,
\end{equation}
for some constant $C$. Applying the boundary condition on the right, we see that $C = V'(\tilde{u}(L'))$. Thus, we obtain a general solution
\begin{equation}\label{eq:gensol}
    \frac{\partial \tilde{u}}{\partial y'} = \sqrt{2} \sqrt{ V'[\tilde{u}(y')] - V'[\tilde{u}(L')] },
\end{equation}
where the choice of positive sign outside the square root is consistent with our boundary conditions.

When $\tilde{u}(0) = d/a$, one solution to Eq. (\ref{eq:gensol}) is $\tilde{u}(y') = d/a$: the signal will propagate completely from one side to the other. The difference in elastic energy from the ground state solution (with $\tilde{u}(y') = -d/a$) will scale with $L$ and, thus, we expect it to be prohibitive for particularly long cascades. However, we also expect solutions with $\tilde{u}(0) = d/a$ but $\tilde{u}(L') < 0$, with a characteristic, dimensionless length $\eta$ governing the penetration of a signal into the chain. Thus we expect solutions, $u(y)$, will transition over a length scale $\ell = \eta a^2 \sqrt{k}/\sqrt{E}$. The elastic energy cost, in this case, will scale with $\ell$ rather than $L$. We thus expect two regimes of behavior: when $\ell < L$, an initial perturbation initiated on the ``left'' of the signaling cascade will penetrate only a finite distance $\ell$; when $\ell > L$, we expect all elements to be on the ``right'' -- the signal will propagate through the entire network. Therefore the equation
\begin{equation}\label{eq:scalingrelationship}
    L = \eta a^2 \sqrt\frac{k}{E}
\end{equation}
represents the boundary between an initial perturbation propagating a finite distance and propagating through the entire network. This equation tells us the scaling relationship between various wire parameters and can be used to predict the general location and shape of the $\ell<L$ and $\ell>L$ regions for a given set of wire parameters.

It is instructive to consider a specific example, for which $V'(\tilde{u}) = (\tilde{u}^2 - d^2/a^2)^2/2$, and Eq. (\ref{eq:overdamped}) takes the form of the generalized Fisher equation \cite{fisher1937wave,kaliappan1984exact}
\begin{equation}
    \frac{\partial u'(y',t)}{\partial t} - D \frac{\partial^2 u'(y',t)}{\partial {y'}^2}+u' \left({u'}^n-1\right)=0
\end{equation}
with $n=2$. Based on solutions presented in \cite{ma1996explicit}, we construct the stationary solution
\begin{equation}\label{eq:usoln}
    u'(y')=-1+2\frac{1}{1+ a e^{\sqrt{2}(y')}}
\end{equation}
valid for a half infinite line starting at $y' = 0$, where $a = (1-u_0)/(1+u_0)$, $\tilde{u}(0) = u_0$.
Putting this back into the elastic energy yields
\begin{equation}
    \mathcal{E} = \frac{(u_0 - 2) (1+u_0)^3}{3 \sqrt{2} },
\end{equation}
for $-1 < u_0 \le 1$. This expression is minimized when $u_0 = -1$, implying that a signal never propagates in a symmetric, infinitely long chain. 
Despite this, finite sized chains are likely to behave differently. Indeed, these results hint that chains with length $L < a^2 \sqrt{k/E}$ -- shorter than the intrinsic scale of Eq. (\ref{eq:usoln}) -- may still propagate signals.

\section{Results}

\subsection{Simulations}

\begin{figure*}
    \centering
    \includegraphics[width=\textwidth]{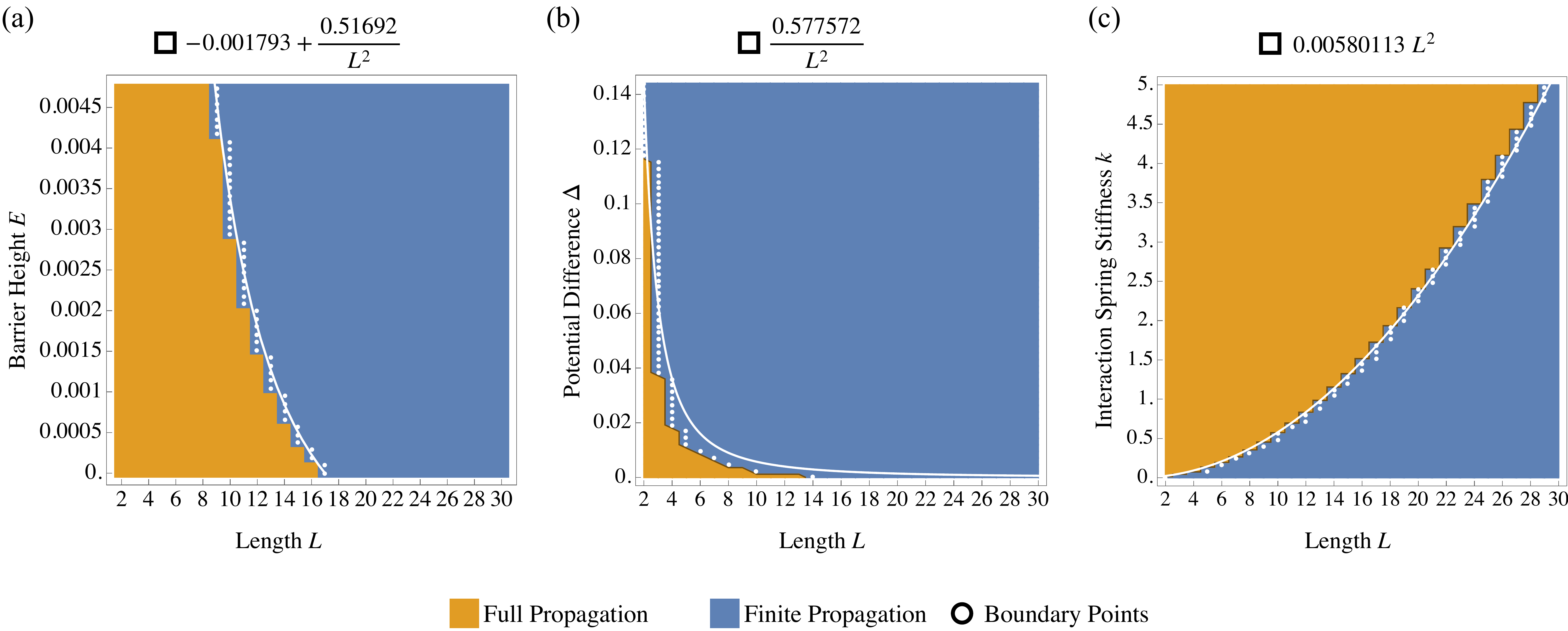}
    \caption{For all three plots, the wire length parameters are $h=1$, $a=1$, and $d=1/4$.(a) Scaling relationship between the potential barrier height $E$ and the length of the wire when varying the beam stiffness. The beam stiffness $b$ ranges from 0 to 5. The interaction spring stiffness $k$ is set to 1, and the torsional modulus $s$ is set to 0. Each boundary point corresponds to a simulated wire with a specific choice of length $L$ and characteristic energy $E$. (b) Scaling relationship between the potential difference between minima $\Delta$ and the length of the wire when varying the torsional modulus. The torsional spring stiffness $s$ ranges from 0 to 0.3. The interaction spring stiffness $k$ and beam stiffness $b$ are both set to 1. Each boundary point corresponds to a simulated wire with a specific choice of length $L$ and torsional modulus $s$. (c) Scaling relationship between the interaction spring stiffness and the length of the wire. The interaction spring stiffness $k$ ranges from 0 to 5. The beam stiffness $b$ is set to 1, and the torsional modulus $s$is set to 0. Each boundary point corresponds to a simulated wire with a specific choice of length $L$ and interaction spring stiffness $k$.}
    \label{simplots}
\end{figure*}

To corroborate our scaling analysis, we performed simulations of the dynamical system in Fig. \ref{chain} directly by integrating Eq. (\ref{eq:EOM}). To obtain the overdamped limit, we set $m = 0$. Consider a wire of length $L$ consisting of bistable elements with equilibrium positions at $x+i=\pm d$. The wire is initially in the ground state with all beams in the $x_i=-d$ ``left" position. To initiate a signaling cascade, we applied a displacement of $2d$ to the first beam to move it into the $x_1=+d$ ``right" position. After a sufficient length of time that the motion has stopped, we recorded the position $x_L$ of the last beam in the wire. If $x_L < 0$, the signal only propagated a finite distance. If $x_L > 0$, the signal propagated through the entire network.

We show the relationship between the barrier height $E$ and the length of the wire $L$ in Fig. \ref{simplots}(a). To verify the scaling relationship between $L$ and $E$ in Eq.(\ref{eq:scalingrelationship}), we fit the boundary between the finite and full propagation regions to an expression of the form $E = C_1 L^{-2} + C_2$, with $C_2 \approx -2 \times 10^{-3}$ and $C_1 \approx 0.51$. This suggests a scaling relationship
\begin{equation}
    E - E_0 = \frac{\eta^2 a^4}{L^2} k
\end{equation}
with $E_0 \approx -2 \times 10^{-3}$. Turning this around,
\begin{equation}
    k = \frac{E-E_0}{\eta^2 a^4} L^2.
\end{equation}

We also show the relationship between the potential difference between minima $\Delta$ and the length of the wire in Fig. \ref{simplots}(b). We find $\Delta \approx C_1 L^{-2} + C_2$, with $C_2 = 0$ and $C_1 \approx 0.58$.

The relationship between $k$ and wire length is depicted in Fig. \ref{simplots}(c) for $E \approx 10^{-3}$. An expression of the form $k \approx C_1 L^2$ with $C_1 \approx 6 \times 10^{-3}$ fits the boundary between the finite and full propagation regions well. Using the previous results for $\eta^2 a^4$, $E$ and $E_0$, we see that $C_1$ is consistent with $(E-E_0)/(\eta^2 a^4) \approx 6 \times 10^{-3}$.

\subsection{Experiments}

\begin{figure}
    \centering
    \includegraphics[width=\columnwidth]{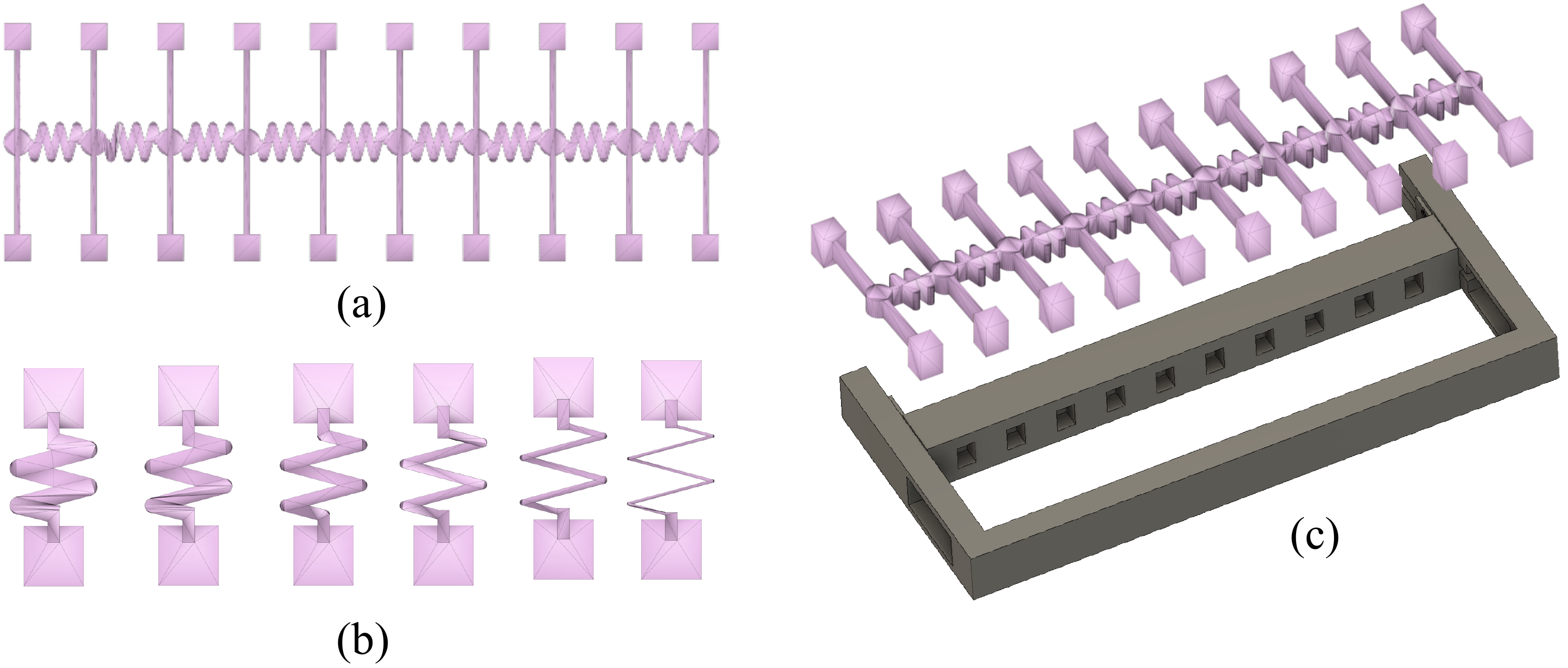}
    \caption{3D models of wire components. (a) The initial state of a wire with symmetric bistable beam elements. (b) Wire slotted into a holding device (frame) that compresses all bistable beam elements to buckle them uniformly. (c) Interaction springs of varying thickness. Different spring stiffnesses are achieved through changing the thickness of the spring. Individual interaction springs are printed specifically for measuring their stiffness.}
    \label{expwiredesign}
\end{figure}

To validate our results experimentally, bistable elements were designed using Fusion 360 CAD software and fabricated using a Formlabs Form2 SLA 3D printer with the Formlabs Elastic 50A resin, which has flexible and stretchable properties after curing. The design for the wires is shown in Fig. \ref{expwiredesign}. The bistable elements were elastic beams that buckle under compression, and the nearest neighbor interactions were facilitated by linear springs. The wire was printed in an unstressed state, and compressed using a rigid frame, printed using Formlabs Grey resin. (Fig. \ref{expwiredesign}).

\begin{figure}
    \centering
    \includegraphics[width=\columnwidth]{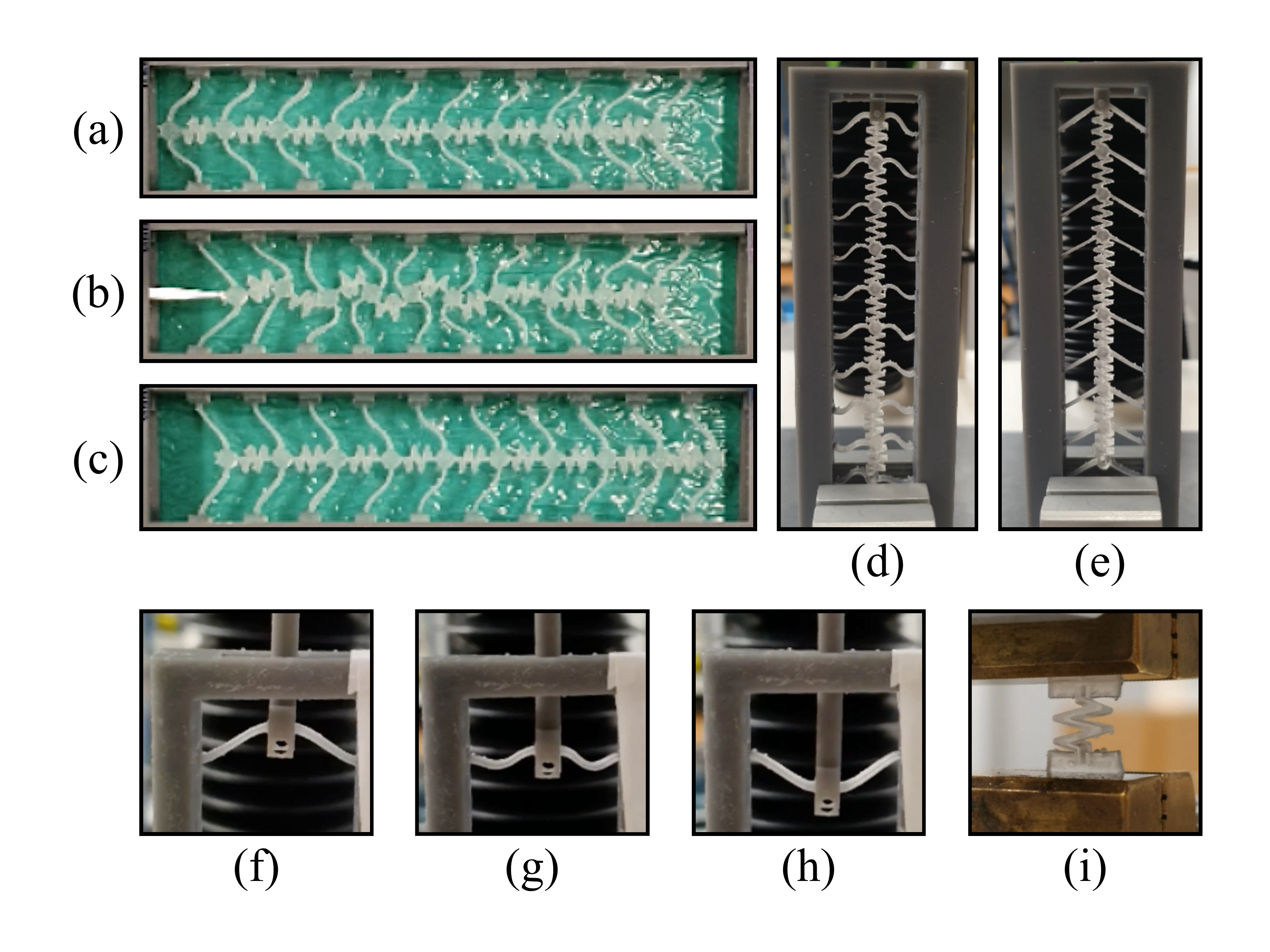}
    \caption{A wire before (a), during (b), and after (c) a signal is fully propagated through manual displacement of the first bistable beam in the wire. Comparison between a wire with symmetric (d) and asymmetric (e) bistable beam elements when mounted in the frame. (f)-(h) Recording the force-displacement data for a single bistable beam. (i) The interaction spring mounted and ready for force-displacement measurements. }
    \label{expwiredata}
\end{figure}

To bias the beams to buckle either to the left or right, pre-curvature was added to the elastic beams by changing the angle of the beam at the fixed end points and midpoint as shown in Fig. \ref{expwiredata}(e). For wires with biased beams, there was no need to compress the wire before sending a signal. 

The wire model used for simulations had seven parameters, $L$, $h$, $d$, $a$, $k$, $b$, and $s$ where values were selected for each. When verifying the scaling relationship found with Eq. (\ref{eq:scalingrelationship}) with simulations, we picked default values for each parameter that simplified our analysis as much as possible. We measured each of these parameters directly from the printed beams in order to run simulations that approximated the behavior of the printed wires. 
The height of the beam $h$ and the distance between beams $a$ were found through direct meaurement of the wire. We determined values for the location of the two potential minima $d$, the beam stiffness $b$, the interaction spring stiffness $k$, and the torsional modulus $s$ using force-displacement measurements of the bistable beams and interaction springs. 

Force-displacement measurements were performed using a custom setup composed of a linear displacement stage (Zaber Technologies Inc., T-LSM 100)
and a load cell (Loadstar Sensors Inc., RPG-10). The wire was placed in the rigid frame and compressed to buckle the beams and set the wire into one of two stable states. The frame was mounted vertically and a signal passed from top to bottom during the test with a mechanical push. Using the linear displacement stage, we slowly pushed on the bistable wire with the load cell and recorded the force exerted on the wire as a function of displacement as the beams transitioned between stable configurations (Fig. \ref{expwiredata}(f-h)). To apply the force to the beam, we used a rigid component printed out of the same material as the frame and attached it directly to the midpoint of the beam. The rigid component was fixed to the wire so that the snap-through transition did not cause the device to lose contact with the load cell.  

We conducted force-displacement measurements for the linear interaction springs using a TA.XTplus texture analyzer (Stable Micro Systems). Each end of the spring was mounted directly in the texture analyzer, as shown in Fig. \ref{expwiredata}(i), and the spring was slowly compressed and stretched. 

To determine the stiffness $k$ of the linear interaction springs, we fit a straight line to the force-displacement data and recorded the slope. To determine an approximate value of $b$ for symmetric beams, we fit the derivative of Eq. \ref{eq:beampotential} to the force-displacement data. For pre-curved beams, we added the torsional spring term Eq. \ref{eq:torsionalpotential} to Eq. \ref{eq:beampotential} before taking the derivative and fitting it to the force-displacement data. Details of the methods used to determine $b$ and $s$ for both symmetric and biased beams are discussed in Appendix A. 

To gather data on what wire parameter combinations allow for finite or full signal propagation, we sent a signal down the wire by hand by pushing on the first beam until it reached the right-buckled position. The beam was held in that position using tweezers for a few seconds to allow the wire to settle into a final position. If the initial displacement caused all subsequent beams to snap through to the right-buckled position, then the signal was considered to have fully propagated along the wire. If the initial displacement moved some of the beams but not all of them, then the signal only propagated a finite distance. By changing the length of the wire and the stiffness of the beams and springs, we then plotted our experimental data the same way we plotted the simulation data in Fig. \ref{simplots} and compared the results. 

\begin{figure*}
    \centering
    \includegraphics[width=\textwidth]{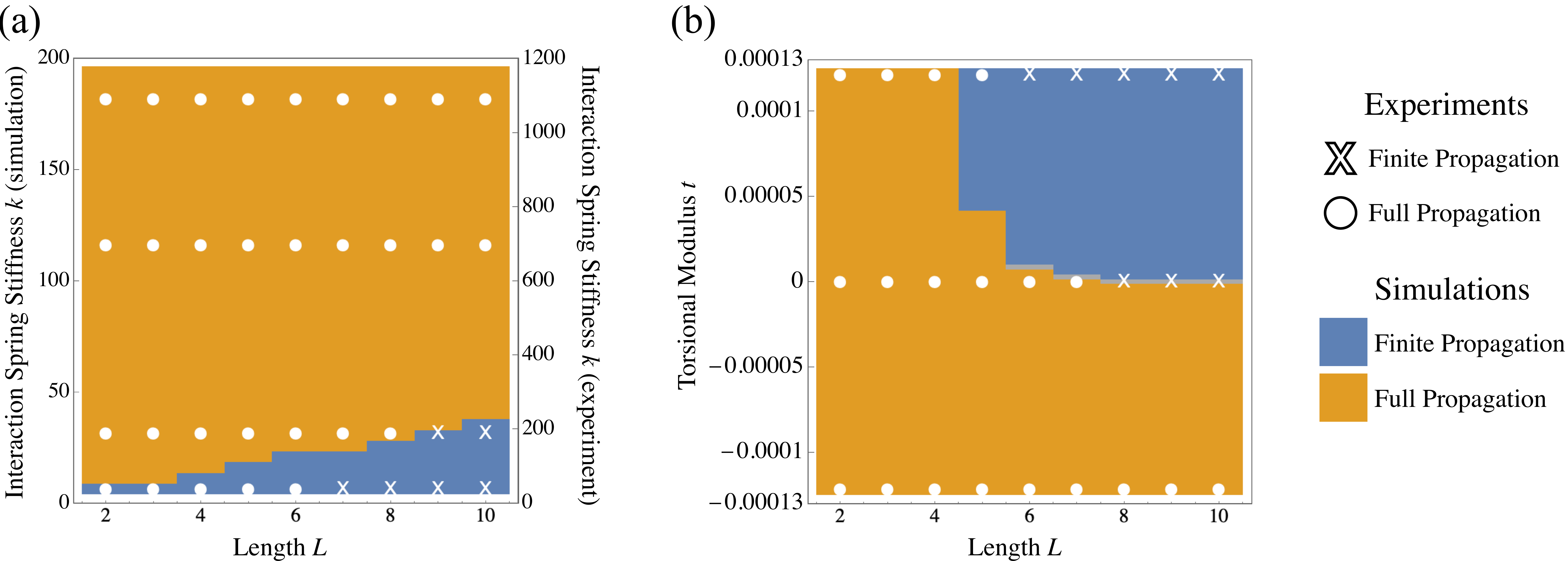}
    \caption{Comparison between signal propagation experiments and numerical simulations. (a) Qualitative comparison between simulations and experiments for varying interaction spring stiffness. The wire parameter values used for simulations are $h=0.012m$, $d=0.0055m$, $a=0.0112m$, $b=120N/m$, and $s=0N/m$.
    (b) Direct comparison between simulations and experiments for varying minima difference $\Delta$ of the bistable beam potential. In simulations, this was done by varying the torsional stiffness on each bistable element. In experiments, this was done by changing the angle of the beam at its fixed endpoints (see Fig. \ref{expwiredata}(e) for example). The wire parameter values used for simulations are $h=0.012m$, $d=0.0043m$, $a=0.0112m$, $b=230N/m$, and $k=30N/m$.}
    \label{simvsexp}
\end{figure*}

Fig. \ref{simvsexp} compares the finite and full propagation regions from experiments and simulations for different values of the interaction spring stiffness and torsional modulus. Comparing the regions produced by varying the interaction spring stiffness shows only a qualitative match between simulations and experiments. When measuring the stiffness of the printed springs, they were only compressed a small amount and were not allowed to buckle side to side. When sending a signal along the printed wires, however, there was substantial buckling of the interaction springs, as shown in Fig. \ref{expwiredata}(b). Thus, the additional buckling lowered the effective barrier height from what was measured. Indeed, we found quantitative agreement between simulation and experimental results if we reduced the simulation interaction spring stiffness by a factor of $6$ from what was initially estimated from deformations that do not buckle.

Using this result, we used a rescaled value for the interaction spring stiffness for the simulations in Fig. \ref{simvsexp}(b). When the torsional modulus was negative, so that propagating a signal causes the beams to transition from a higher energy state to a lower energy state, we confirmed that signals always propagate fully for any number of elements. For positive values of the torional modulus, however, signal propagation depended on the number of elements and was consistent with predictions from simulations. 

\section{Conclusion/Discussion}

In this paper we analyze signaling cascades of bi-stable mechanical elements coupled by springs triggered by the imposition of a fixed displacement on one of the elements. While signals can propagate at any distance when the bistable elements switch from a higher energy to lower energy state, here we find that signals can only propagate a finite distance when the elements transition to either a higher energy state or a state with the same energy. Numerical analysis of simulation data in the overdamped limit shows that signals can propagate when
\begin{equation}
    L < \eta a^2 \sqrt{\frac{k}{E-E_0}},
\end{equation}
even for a completely symmetric signalling cascade.

Though our analysis was for a simple signalling cascade, one expects even more complex digital mechanical logic elements to also present energy barriers that must be overcome by a propagating signal. For true reversible logic, this finite propagation length might present a true limitation to the size and complexity of a device when realizing repeatable, complex logic in a soft mechanical system. 

\begin{acknowledgments}
The authors gratefully acknowledge support for this work provided by the US Army Research Office through grant W911NF-21-1-0068.
\end{acknowledgments}

\bibliography{signalprop}

\appendix

\section{Fit for Simulation Parameters}

\subsection{Interaction Springs}

We fit a straight line that crosses the origin to the force-displacement data for springs of varying thickness. The slope of the line is the linear spring stiffness of the printed springs. 

\begin{table}[h]
    \caption{Interaction spring stiffness for various beam thicknesses.}
    \begin{ruledtabular}
        \begin{tabular}{c c}
            Thickness(mm) & Stiffness (N/m) \\
            \colrule
            0.6 & 30.77 \\ 
            0.8 & 187.12 \\ 
            1.0 & 715.47\\ 
            1.2 & 1123.95 \\ 
        \end{tabular}
    \end{ruledtabular}
\end{table}

\subsection{Symmetric Bistable Beams}

We recorded force-displacement data for a single bistable beam and integrated it numerically using a custom MATLAB program to get the potential-displacement data for our printed beams. We measured $h$ directly from the wire. We calculated $d$ by taking half the distance between the two minima in the potential-displacement data. 

We then took the beam potential $V(x)$ from Eq. (\ref{eq:beampotential}) and fit it to our force- and potential-displacement data in one of five ways: fit $-dV/dx$ to the maximum of the force data, fit $-dV/dx$ to the minimum of the force data, fit $V(x)$ to the energy barrier of the potential data, or by using the built-in Mathematica function FindFit with both the force and potential data. The first three values of $b$ are averaged together to give an approximate beam stiffness, and the last two are left as-is. 

\begin{table}[h]
    \caption{Dimensions for symmetric bistable beams}
    \begin{ruledtabular}
        \begin{tabular}{c c c}
            End-to-End Distance (mm) & h (m) & d (m) \\ 
            \colrule
            22 & 0.011 & 0.0066 \\
            24 & 0.012 & 0.0055 \\
            26 & 0.013 & 0.0043 \\
        \end{tabular}
    \end{ruledtabular}
\end{table}

\begin{table}[h]
    \caption{Beam stiffness for symmetric bistable beams.}
    \begin{ruledtabular}
        \begin{tabular}{c c c c}
            & & \multicolumn{2}{c}{FindFit Stiffness (N/m)} \\     
            \multicolumn{1}{c}{\parbox[t]{2.4cm}{End-to-End\\Distance (mm)}} & \parbox[t]{2.4cm}{Average\\Stiffness (N/m)} & \parbox[t]{1.5cm}{Force Data} & \parbox[t]{1.5cm}{Potential Data}\\ 
            \colrule
            22 & 80 & 83 & 83 \\ 
            24 & 119 & 135 & 123 \\ 
            26 & 187 & 211 & 178\\ 
        \end{tabular}
    \end{ruledtabular}
\end{table}

\subsection{Asymmetric Bistable Beams}

For the bistable beams printed with pre-curvature, we need to determine the torsional modulus $s$ and the beam stiffness $b$. Just as with the symmetric beams, we measured $h$ directly from the wire and calculated $d$ by taking half the distance between the two minima in the potential-displacement data. We took the beam potential $V(x)$ that included the terms in both Eq. (\ref{eq:beampotential}) and (\ref{eq:torsionalpotential}) and used the built-in Mathematica function FindFit to solve for both $b$ and $s$ at the same time. 

\begin{table}[h]
    \caption{Beam parameters for asymmetric bistable beams. First row: the beam transitions from a lower minimum to a higher minimum (uphill push). Second row: the beam transitions from a higher minimum to a lower minimum (downhill push). }
    \begin{ruledtabular}
        \begin{tabular}{c c c c c}
            & h (m) & d (m) & b (N/m) & s (J) \\ 
            \colrule
            \multicolumn{1}{c}{Uphill Push} & 0.012 & 0.0043 & 233 & 0.00013 \\
            \multicolumn{1}{c}{Downhill Push} & 0.012 & 0.0047 & 162 & 0.00013 \\
            \end{tabular}
    \end{ruledtabular}
\end{table}

\end{document}